# Thermally-driven formation of Ge quantum dots on self-catalysed thin GaAs nanowires

Yunyan Zhang [†,¶,*], H. Aruni Fonseka[‡], Hui Yang[#], Xuezhe Yu[†], Pamela Jurczak[†], Suguo Huo[§], Ana M. Sanchez[‡] & Huiyun Liu[†]

[†]. Department of Electronic and Electrical Engineering, University College London, London WC1E 7JE, United Kingdom

[‡]. Department of Physics, University of Warwick, Coventry CV4 7AL, United Kingdom

[#]. Department of Materials, Imperial College London, Exhibition Road, London SW7 2AZ, United Kingdom

[§]. London Centre for Nanotechnology, University College, London WC1H 0AH, United Kingdom

[¶]. Department of Physics, Paderborn University, Warburger Straße 100, 33098, Paderborn, Germany

**Abstract:**

Embedding quantum dots (QDs) on nanowire (NW) sidewalls allows the integration of multi-layers of QDs into the active region of radial p-i-n junctions to greatly enhance light emission/absorption. However, the surface curvature makes the growth much more challenging compared with growths on thin-films, particularly on NWs with small diameters (Ø <100 nm). Moreover, the {110} sidewall facets of self-catalyzed NWs favor two-dimensional growth (2D), with the realization of three-dimensional (3D) Stranski-Krastanow growth becoming extremely challenging. Here, we demonstrate thermally-driven formation of Ge dots on the {110} sidewalls facets of thin self-catalyzed NWs without using any surfactant or surface treatment. The 2D-3D transition of the pseudomorphic Ge layer grown on GaAs NWs is driven by energy minimization under high-temperature annealing. This method opens a new avenue to integrate QDs on NWs without any restriction on NW diameter or elastic strain, which can allow the formation of QDs in a wider range of materials systems where the growth of islands by traditional mechanisms is not possible, with benefits for novel NWQD-based optoelectronic devices.

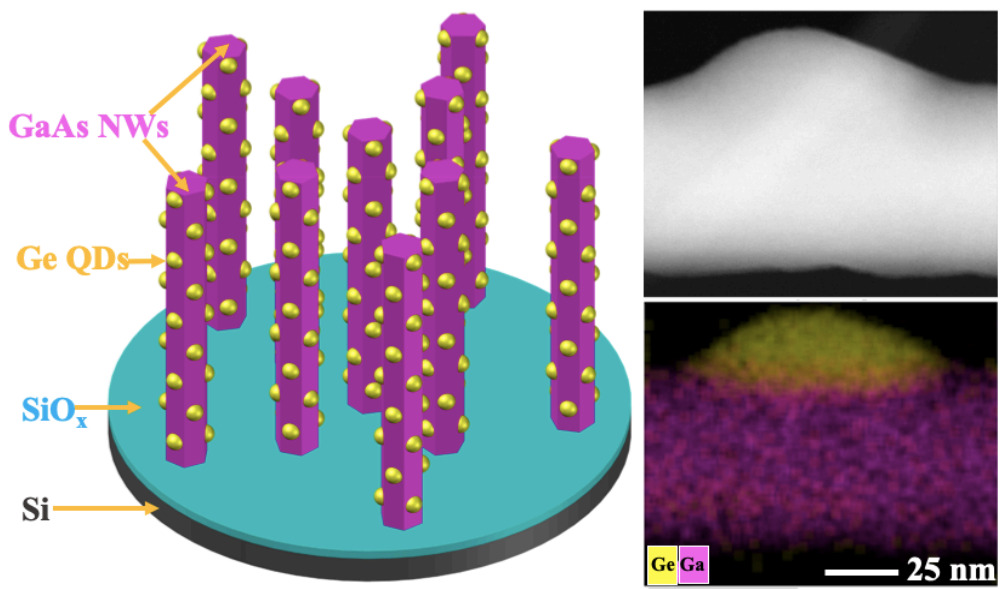

**Introduction**

Nanowires (NWs) have novel mechanical, optical and electronic characteristics that are not observed in thin films, for example operating as an optical cavity for light emitters and enhancing light absorption for photovaltaics.[1–5] Hybrid structures, designed by integrating quantum dots (QDs) with NWs, further improves their performance compared with homogeneous NWs. QDs provide a unique semiconductor structure with fully quantised electronic states for the fabrication of both high-efficiency classical and non-classical photon emitters/absorbers. Examples include low threshold current density lasers with reduced sensitivity to temperature variation, trains of regularly spaced single photon emitters, and entangled photon pair sources.[6-8]

Research on QD/NW integration has gained increasing attention in recent years. Especially, the ability to grow QDs on NW sidewalls can allow easy control of the thickness of shells beneath and above the QD layers, which is advantageous for sandwiching QDs into the intrinsic active region of radial p-i-n junctions. Besides, many optoelectronic devices such as lasers and intermediate band solar cells require multilayer QDs to obtain higher gain or light absorption.[9] This can be achieved by growing multilayer QDs on the NW sidewalls, with many reports available on the realization of this type of structures, such as Ge QDs on Si NWs, MnAs QDs on InAs QDs, and In(Ga)As QDs on GaAs NWs.[10-14] A majority of these studies are based on the strain-driven Stranski-Krastanov (SK) growth mode[11-14,15], although dot formation at the junction of branched NWs[16] and as a result of phase separation[17] have also been observed. SK growth mode is responsible for the 2D to 3D transition in heteroepitaxial growth and it has been successfully applied in low-threshold lasers, light emitting diodes, and solar cells.[6,18] However, the implementation of SK growth mode on NWs faces significant challenges.

The SK growth mechanism occurs to relieve the strain energy of lattice mismatched hetero-epitaxy through the formation of island. Thus, for example in the InAs/GaAs system, with 7.2% lattice mismatch between InAs and GaAs, InAs QDs are formed as follows. Initially, a complete two-dimensional InAs pseudomorphic wetting layer (WL) on the GaAs surface forms, followed by the growth of three dimensional InAs islands. The WL thickness depends on the nature of the system, thus above a critical thickness of the WL, the strain energy stored in the epilayer causes the surface morphology to change from 2-dimensional (2D) layer growth to an island mode (3D).

In the case of epitaxial growth on NWs, the NW can share strain caused by the lattice mismatch of the QD on its facets when the NW thickness is similar in size to the QD width.[19,20] Therefore, QD growth on NWs requires a much thicker critical wetting layer thickness for island formation than that on a planar substrate.[21,22] The presence of a thick WL interconnecting QDs on thinner wires can greatly degrade QD performance. It has been found that QDs form more readily on a thicker NW in comparison with a thinner one, because less material contributes to the wetting layer shell and more contributes to the formation of QDs.[23] For very thin NWs, the growth may only follow the Frank–van der Merwe (FM) thin-film growth mechanism due to the good radial strain relaxation ability, which completely prevents the growth of QDs.[24,25] It has been proposed that the critical diameter of a GaAs NW to form InAs QDs is larger than 120 nm.[26]

Moreover, the SK growth mode is very sensitive to surface orientation. Gold-catalysed NWs exhibit {112} sidewall facets, which are generally favourable for the SK QD growth. However, these NWs are incompatible with Si-based electronics, as Au can be incorporated into Si substrate[27,28] as well as NWs[29,30] at levels of the order of $10^{15}$–$10^{18}$ cm$^{-3}$. In the case of Au-free NWs, such as the self-catalysed NWs, {110} sidewall facets are commonly observed in the NWs.[31] Due to the non-favourable {110} surface, QDs do not form on these surfaces, and micrometer-sized triangular structures containing dislocations are commonly observed.[32-35] To the best of our knowledge, there is just one report on successful QDs growth on {110} NWs sidewalls so far. This was achieved by modifying the NW surface using an AlAs layer prior to QDs growth.[36] A direct growth technique without complex surface treatment is preferred to simplify the growth procedure and avoid the degradation of QD performance by additional layers.

On the other hand, a NW's ability to facilitate defect-free integration of different material families can be used to combine material systems that are highly advantageous for enhancing the performance of existing devices, and also creating new applications.[37] Especially, the integration between the two most powerful semiconductor groups, group IVs and group III-Vs, is particularly relevant for expanding the wealth of optical and electronic applications.[38,39]

In this article, we report the formation of Ge quantum dots on self-catalysed thin GaAs nanowires by a thermally-driven mechanism, providing a novel approach to form high-quality QDs on the {110} NW sidewalls regardless of the NW size and free from the formation of wetting layers.

**Results and discussion**

*Influence of temperature on Ge growth*

GaAs core NWs with a diameter ~50 nm were grown by self-catalyzed mode on Si substrates. These NWs exhibited dominantly zinc blende crystal structure and smooth {110} sidewalls dominating this NWs. Single twins were presents just at the very tip and bottom parts (see Reference 40 and 41 for more details on these same GaAs core NWs). A Ge shell, with a nominal thickness of ~50 nm (calibrated according to thin-film growth rate), was then grown on the GaAs core NWs at three different temperatures. The morphology and crystal quality of these Ge/GaAs NWs were analysed by Scanning Electron Microscopy (SEM) and Transmission Electron Microscopy (TEM) imaging respectively (Figure 1). When the Ge shell is grown at 450 ℃, the {110} NW facets parallel to the growth direction form flat sidewalls (Figure 1a). The occasional twin defects (indicated by red arrows in Figure 1d) commonly observed at the tip and bottom of the GaAs core NW has no apparent influence on the morphology of the {110} sidewalls (Figure 1d). The increase on Ge growth temperature to 500 ℃ results in nano-scale sidewalls roughness, giving a wavy appearance (Figure 1b), as clearly observed in Figure 1e. Further increase of the growth temperature up to 550 ℃, causes the surface to roughen even further and large lump-like features are form on NW sidewall surfaces (Figure 1c and f). Energy-dispersive X-ray spectroscopy (EDX) was performed on these NWs to ascertain the elemental distribution. Figure 1g correspond to the annular dark field image and As, Ga, and Ge elemental distribution in a NW with the Ge shell grown at 550 ℃. As can be observed, both Ga and As maps agrees with a smooth sidewall of the GaAs core NWs, whilst the Ge elemental distribution corroborates that the sidewall roughness is mainly caused by the Ge shell grown at 550 ℃.

Large lattice mismatch between different materials results in structure defects and lattice distortion. Misfit accommodation can be also accommodated by the introduction of island or surface roughness, commonly associated to the large lumps. Since the strain energy stored in mismatched layer increases usually with the square of the film thickness , and QDs formation is often considered as a form of strain relief. Thus, in SK growth mode, the formation of QD is a strain-driven process during which a 2D-3D transition occurs, with reorganization of the growing surface. In our system, the lattice mismatch between Ge and GaAs is <0.1%, ruling out strain-driven dots formation or surface roughness as revealed in Figure 1f. Another factor that can lead to surface roughness is the presence of planar defects such as twins and stacking faults.[42] Nevertheless, the lump-like features also emerge in NWs segments which are stacking fault free, as presented in Figure 1h and further confirmed by the Selected Area Electron Diffraction (SAED) pattern in Figure 1i.

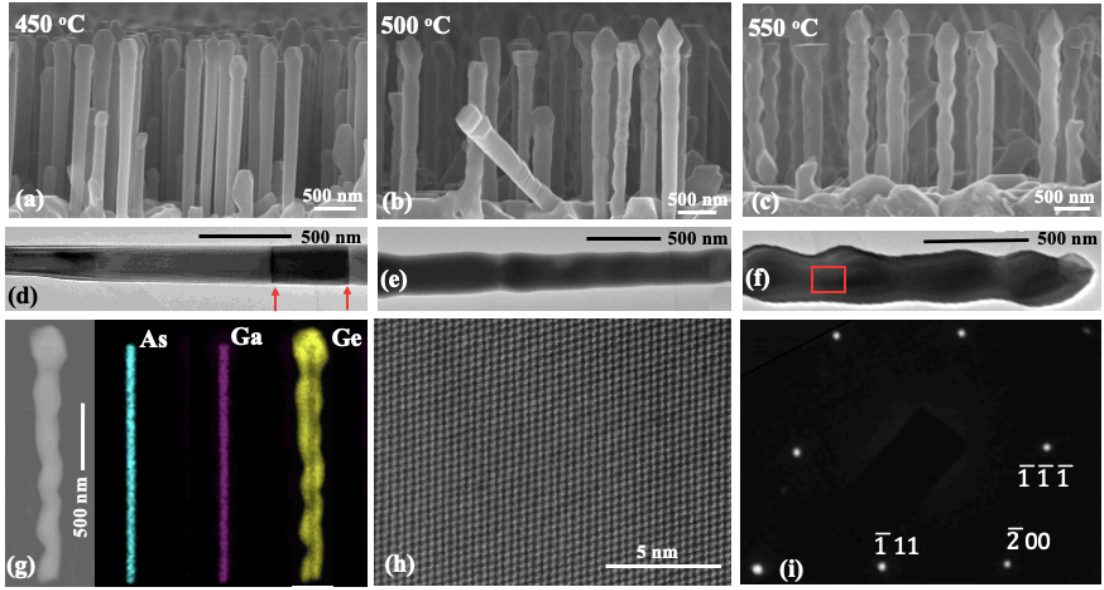

**Figure 1. GaAs-Ge core-shell NWs. SEM and TEM images of the NWs with Ge shell.** Ge shell grown at **a, d** 450 ℃, **b, e** 500 ℃, and **c, f** 550 ℃. EDX maps of **g** a GaAs/Ge NW from (c) showing the distribution of As (cyan), Ga (magenta), and Ge (yellow) elements. **h** A representative atomic-resolution STEM image and **i** SAED pattern of the regions marked by a red square in (f).

**Thermodynamics of the rough Ge shell growth**

To elucidate the thermally-driven nature of the surface roughness and lump-like features observed in our system, i.e. Ge/GaAs NWs, theoretical calculations were performed using plane-wave pesudopotential density functional theory (DFT) formalism implemented in Vienna Ab-initio Simulation Package (VASP)[43]. The structural stability is characterized by the crystal binding energy (ΔE), i.e. the system energy change before and after the Ge-GaAs integration, defined as:

$$\Delta E = E_{GaAs/2Ge} - E_{GaAs} - 2 E_{Ge}$$

Where, $E_{GaAs}$, $E_{Ge}$ and $E_{GaAs/2Ge}$ are the total free energies of GaAs, Ge and GaAs-2Ge unit cells, respectively. GaAs and Ge crystal have the same zinc combined structure with nearly the same lattice constant of 5.65 Å and 5.66 Å, respectively. Therefore, in the Ge/GaAs blender system, Ge atoms directly sit on the Ga/As equivalent sites, so that one portion Ga atoms and one portion of As atoms atom are replaced by two portion of Ge atoms. Figure 2 illustrates the crystal structures of the three different cells, GaAs, Ge, and their combination on (110) facets. The results revealed that the system energy increased by 8.28 meV/Å$^2$ when GaAs and Ge are integrated, suggesting that the 2-dimensional Ge overlayer on GaAs is at a higher energy than a Ge overlayer on bulk Ge due to the incoherence of the electronic states.

At a low temperature of 450 ℃, 2-dimensional Ge thin-film growth (pseudomorphic mode) occurs under non-thermodynamic equilibrium conditions. Here, the Ge shell grows in a quasi-FM mode and the morphology is determined by a random but uniform deposition. When the growth temperature increases, so does the ad-atom mobility, enabling the system to reach an energetically more stable state. The atoms of the Ge film are more strongly coupled with each other than with the GaAs. This leads to 3D islands nucleation and growth directly on the surface of the NW, lowering the total system energy by allowing excess Ge to exist in an electronic state similar to bulk Ge. Thus, the substrate (i.e. NW) influences the overlayer not through strain but rather electronically.[44]

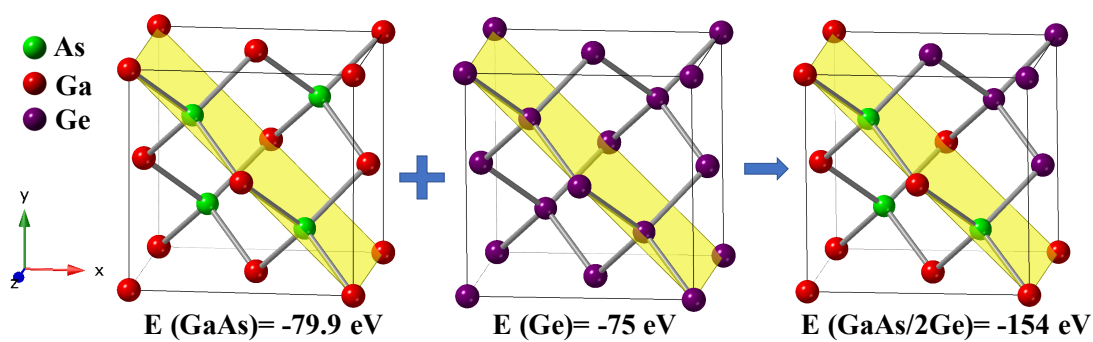

**Figure 2. The crystal structure illustration of bulk GaAs, Ge and integrated GaAs/2Ge.** The (110) lattice plane is showed in yellow. The total free energies are labelled per computational cell and the combined system is higher in energy by 8.28 meV/Å$^2$.

**Ge dot formation on NW sidewalls**

As experimentally and theoretically demonstrated above, thermally-driven surface corrugation occurs in Ge/GaAs NWs. This phenomenon can provide a successful route to generate three-dimensional island on self-catalyzed NWs sidewalls. This is a very promising process to solve long-term difficulties in growing QDs on self-catalyzed small NWs. To demonstrate this, a Ge shell with a nominal thickness of ~10 nm was grown on GaAs core NWs. The growth temperature was 350 ℃, beneficial for uniform shell layer formation. Then, the NWs was annealed at 600 ℃ for 15 mins.

As can be seen in Figure 3a and b, the GaAs NW surface is decorated with strings of dots from bottom to tip. These dots are preferentially located on the sidewall facets rather than on the corners, and the dot size is larger at the locations where the facet is wider, which is shown in Figure 3c. If the side facet width is uniform along the NW length, the dots are relatively uniform in size and equally-distanced as can be seen in Figure 3d. When the NW width is too large, it forms multiple dots (Figure 3e). In thinner wires, the dots are spanning over to the two facets on the sides (Figure 3f), which could be avoided by reducing the Ge deposition thickness. The EDX elemental maps demonstrate that the dots are mainly Ge in composition (Figure 3 g, h and i). A more detailed image and compositional analysis is shown in Figure 3j and k. Here, it is clearly observed that Ge agglomerates to form dots on the NW sidewalls. Compositional line profile across the NWs in Figure 3l shows that only the dot region of NWs has the detectable Ge; while the rest has the Ge signal at background level.

The high temperature annealing enhanced the Ge atoms mobility, beneficial for the minimization of system energy and gathering the Ge atoms into dots formation. The preferred location on the facet planes rather than the corner may be related to the small contact area of the dots with NWs and hence have a lower system energy. The increase in the dot size with the facet width is due to the larger facet being able to provide a larger amount of material for the formation of dots.

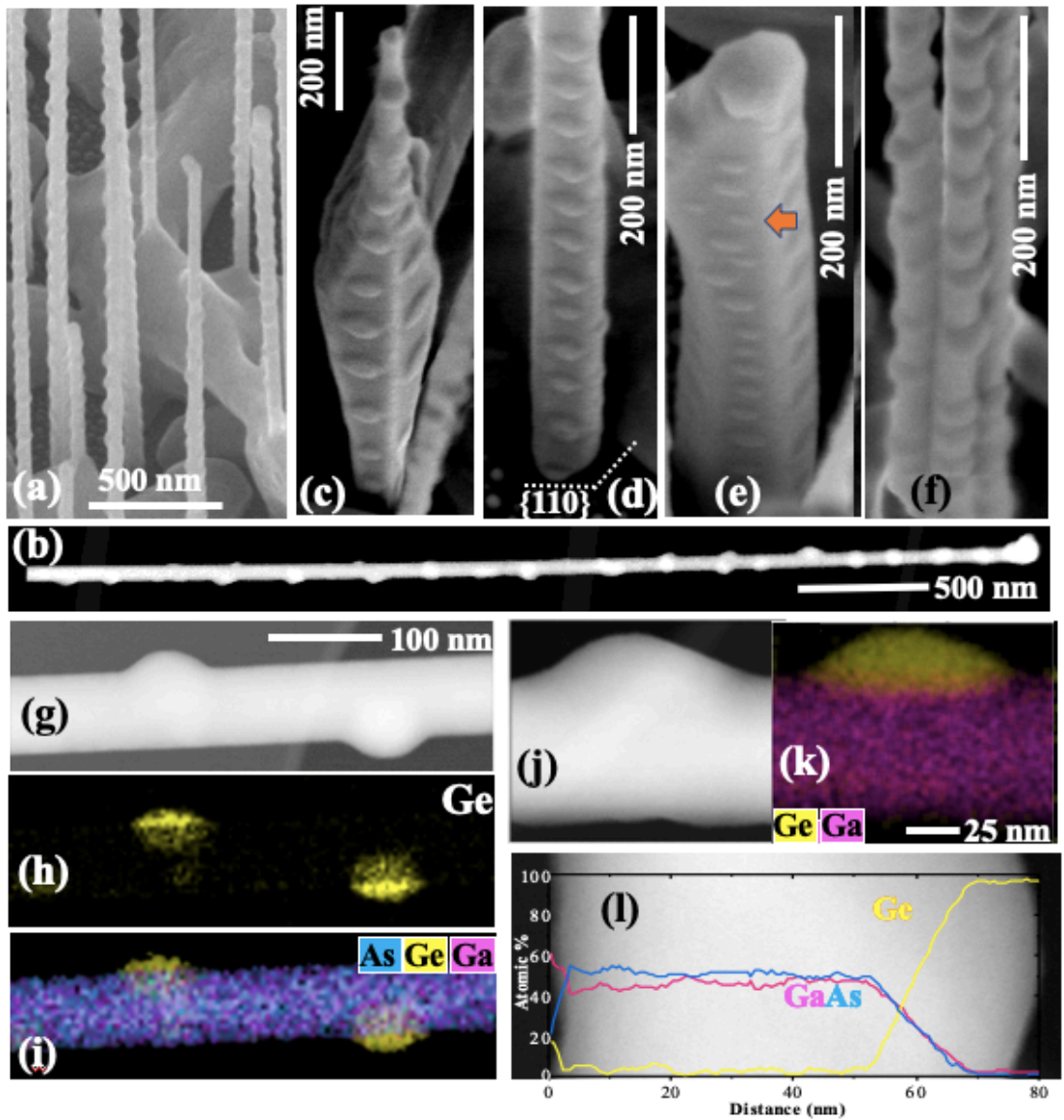

**Figure 3. Ge dots grown on GaAs NWs. a, c, d, e** and **f** are SEM images of the NWs with different facet sizes. Low magnification TEM shows **b** the entire NW and **g** a NW segment. EDX mapping of **h** Ge and **i** composition overlay of the NW segment shown in (g). **j** TEM and **k** corresponding EDX mapping of a Ge dot. **l** EDX line profile across the NW at a Ge dot.

**Crystalline properties of Ge dots**

Ge dots formed during the annealing process are epitaxial with respect to the core NWs as shown in Figure 4a and b. The Ge dots atomic arrangement follows the zinc blende crystal structure observed in the GaAs. No dislocation was observed in the Ge dot or at the dots/core NW interface, confirmed by the strain map of the interface region shown in Figure 4c. The strain map analysis using the geometric phase algorithm did not expose any detectable strain.[45] When grown on a twin plane region (Figure 4d), the dots follow completely the core NW crystal template, resulting in Ge dots contain also twin planes (Figure 4e), without having any influence on their shape. We have not found any correlation between the dot formation and the presence of stacking faults, i.e. the defects did not act as preference site on the dot formation. The dots can be grown on twin planes (Figure 4e), but without specific preference to grow on them. As can be seen in Figure 3f and g, the dots are not pinned to twin planes despite being only a few nanometres away.

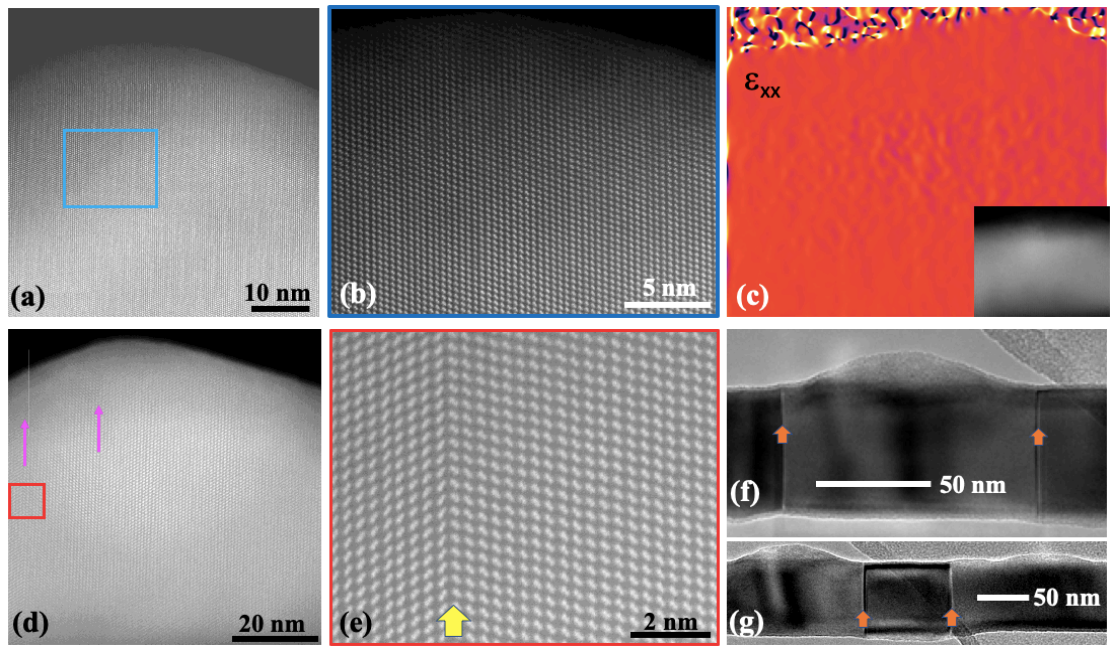

**Figure 4. TEM images of Ge dots. a** low-resolution image of a defect-free dot. **b** Atomic resolution image of the region shown in the blue region of (a). **c** Strain map in the horizontal direction of a dot. **d** A low magnification image of a dot containing two single twins. **e** Atomic resolution image of the region shown in (d). Low magnification image of dots grown **f** between and **g** next to two single twins.

In conclusion, we have demonstrated for the first time the growth of strain-free dots on the {110} sidewalls of self-catalyzed thin NWs. The Ge dot formation is thermally-driven to minimize the system energy, which has much less restrictions on the requirement on the facet size than the SK growth mode. The dots can have relatively uniform size and inter-dot distance when grown on facets with a uniform width along the NW length. Benefited from this growth mode, the dot formation does not need the assistance of the wetting layer that commonly seen in SK dots, which can provide better quantum confinement. There is no observation of defects in these dots or at the dot/NW interface, and there is no direct correlation between the dot growth and the stacking faults of the core NW. This study solved the long-term issue of growing high-quality QDs onto the sidewall of NWs, and the strain-free growth mode can allow the construction of QDs from a much wider range of materials than traditional SK growth mode, allowing the development of novel QD devices. Besides, this also demonstrates the integration of high-quality group-IV materials with self-catalyzed III-V NWs in the form of QDs, which is promising for the construction of novel optical and electronic devices.

**Experimental section**

*NW growth*: The self-catalyzed GaAs NWs were grown directly on p-type Si substrates by means of solid-source III−V molecular beam epitaxy (MBE).[46] GaAs NWs were grown with a Ga beam equivalent pressure, V/III flux ratio, substrate temperature, growth duration and Be flux (equivalent to a nominal thin-film doping concentration) of $8.41\times10^{-8}$ Torr, 44, ~630°C, 1 hour and $0~1.28\times10^{19}/cm^3$, respectively. After cooling down, the samples were transferred to the solid-source group-IV MBE for Ge growth using the ultrahigh-vacuum transfer chamber connecting the two MBE systems, which can keep a minimum contamination and surface damage of GaAs NWs when without exposing samples to air. The thick Ge shell (shown in Figure 1) was grown at different temperatures (450, 500, and 550 °C) with a flux rate of $1.57 \times 10^{-7}$ Torr for 1 h. The Ge QDs was grown by depositing 10 nm Ge and then annealed at 600 °C for 15 mins. The thin Si capping was grown at 550 °C.

*Density functional theory modelling*: The starting points for our calculations were the experimentally determined structures of GaAs[47] and Ge[48] shown in Figure 2. The model cells have been generated by projecting their primitive cubic cell on the lattice plane with Miller indices (h=1, k=1, l=0) to accomplish combining the two-parental crystal. The crystal transformation and the images were generated using CrystalMaker[49]. As a result, the number of atoms (see table 1) are 16, 16 and 32 atoms in the lattice energy minimization simulations for GaAs, Ge and GaAs/2Ge, respectively. The blender interface is in *ab* plane in the transformed system.

Table 1. lattice constants, number of atoms and free energy of transformed unit cell for GaAs, Ge and GaAs/2Ge crystals

|  | No of atom/unit cell | *a* | *b* | *c* | *Free energy (eV)* |
|---|---|---|---|---|---|
| Ge | 16 | 5.76 | 8.15 | 8.15 | -79.90 |
| GaAs | 16 | 5.75 | 8.13 | 8.13 | -75.00 |
| GaAsGe | 32 | 5.75 | 8.13 | 16.26 | -154.00 |

For DFT calculations, Perdew-Burke-Ernzerhof (PBE) functional[50] was used alongside dispersion correction of the Grimme's method (DFT-D3) [51] for geometry optimization. Γ-centered Monkhorst−Pack k-point meshes with $3 \times 2 \times 2$ subdivisions were used for Ge and GaAs, while a $3 \times 2 \times 1$ mesh was used for GaAs/Ge$_2$ system because the combined crystal has a more anisotropic unit cell. The energy cutoff for the plane-wave basis set was set to 600 eV, and projector augmented-wave pseudopotentials[52] were used. During the geometric optimization, the total energy was converged to within 1 meV per atom, and a force tolerance of 0.001 eV/Å was set for convergence of the ion positions to ensure accurate atomic positions and lattice parameters.

***Scanning electron microscope (SEM)***: The NW morphology was measured using Zeiss XB 1540 FIB/SEM and Zeiss Gemini 500 SEM systems.

***Transmission electron microscopy (TEM)***: Simple mechanical transfer of the NWs onto lacey carbon support grids was used to prepare TEM specimens. The TEM measurements were performed using doubly−corrected ARM200F and Jeol 2100 microscopes, operating at 200 kV. EDX measurements were performed using an Oxford instruments 100 mm$^2$ windowless detector installed within the ARM200F microscope.

**ASSOCIATED CONTENT**

**Author Information:**

Corresponding Author: Yunyan Zhang

*E-mail: yunyan.zhang.11@ucl.ac.uk

**Declaration of Competing Interest:**

The authors declare no competing financial interest.

**Acknowledgements:**

The authors acknowledge the support of Leverhulme Trust, EPSRC (grant nos. EP/P000916/1 and EP/P000886/1), and EPSRC National Epitaxy Facility.